\documentclass[aps,twocolumn,preprintnumbers,amsmath,nofootinbib,amssymb]{revtex4-2}

\usepackage{graphics}
\usepackage{epsfig,epsf,psfrag}
\usepackage{times}
\usepackage{float}
\usepackage{color}
\usepackage{amsfonts,amssymb,stmaryrd,latexsym,amsmath}
\usepackage{multirow}
\usepackage{array}
\usepackage{hhline}
\usepackage{booktabs}
\usepackage{enumerate}
\usepackage{slashed}
\usepackage{subfigure}
\usepackage[colorlinks, citecolor=blue,anchorcolor=red,menucolor=red, linkcolor=red,filecolor=red,runcolor=red,urlcolor=blue,frenchlinks=red]{hyperref}
\usepackage{epstopdf}
\usepackage{braket}

\allowdisplaybreaks[4]

\begin{document}

\title{Identifying the two-pole structure of the $\Lambda(1405)$ using  an SU(3) flavor filter}

\author{Ying-Bo He$^{1}$}
\author{Xiao-Hai~Liu$^{1}$}~\email{xiaohai.liu@tju.edu.cn}
\author{Li-Sheng~Geng$^{2,7,8}$}~\email{lisheng.geng@buaa.edu.cn}
\author{Feng-Kun Guo$^{3,4,7,8}$}~\email{fkguo@itp.ac.cn}
\author{Ju-Jun~Xie$^{5,6,7}$}~\email{xiejujun@impcas.ac.cn}

\affiliation{ $^{1}$Center for Joint Quantum Studies and Department of Physics, School of Science, Tianjin University, Tianjin 300350, China\\
$^{2}$School of Physics, Beihang University, Beijing 102206, China\\
$^{3}$CAS Key Laboratory of Theoretical Physics, Institute of Theoretical Physics, Chinese Academy of Sciences, Beijing 100190, China\\
$^{4}$School of Physical Sciences, University of Chinese Academy of Sciences, Beijing 100049, China\\
$^{5}$Institute of Modern Physics, Chinese Academy of Sciences, Lanzhou 730000, China\\
$^{6}$School of Nuclear Sciences and Technology, University of Chinese Academy of Sciences, Beijing 101408, China\\
$^{7}$ Southern Center for Nuclear-Science Theory (SCNT), Institute of Modern Physics, Chinese Academy of Sciences, Huizhou 516000, China\\
$^8$ Peng Huanwu Collaborative Center for Research and Education, Beihang University, Beijing 100191, China}	

\date{\today}
	
\begin{abstract}
We propose a novel method to identify the two-pole structure of the $\Lambda(1405)$. The two poles owe their origin to different quark flavor irreducible representations in the meson-baryon coupled-channel interactions, thus they should be individually manifested in reactions that provide good flavor eigenstate sources. Hadronic decays of charmonia into $\bar{\Lambda}\Sigma\pi$ and $\bar{\Lambda}(1520)\Sigma\pi$ are such reactions, and the flavor octet  and singlet poles can be approximately singled out in these two decay modes. 
This SU(3) flavor filter  works even considering the flavor symmetry breaking. With the huge charmonium data sets collected, it is therefore promising to solve the long-standing $\Lambda(1405)$ puzzle employing the proposed flavor filter.

\end{abstract}

\maketitle

{\it Introduction.} The $\Lambda(1405)$ is a hyperon resonance with quantum numbers $J^P=1/2^-$ and $I=0$. It was first predicted as a $\bar K N$ bound state by Dalitz and Tuan~\cite{Dalitz:1959dn} before it was observed in 1961~\cite{Alston:1961zzd}. Although it is currently quoted by the Particle Data Group (PDG)~\cite{ParticleDataGroup:2022pth} as a four-star resonance, its underlying structure is still under debate. In the constituent quark model, it is difficult to assign the $\Lambda(1405)$ as the first orbital excited state of the $\Lambda$ hyperon, because its mass is much lower than its nucleon counterpart $N(1535)$, and the mass 
splitting between the $\Lambda(1405)$ and its spin partner $\Lambda(1520)$ with $J^P=3/2^-$ is much larger than that in the nucleon sector. 
Thus, it has been a long-standing puzzle what the internal structure the $\Lambda(1405)$ possesses.
The modern theory concerning the $\Lambda(1405)$ is the coupled-channel 
theory constrained by chiral and SU(3) flavor symmetry as well as  unitarity~\cite{Kaiser:1995eg,Oset:1997it,Lutz:2001yb,Oller:2000fj,Oset:2001cn}. In this theory, the $\Lambda(1405)$ is interpreted as a dynamically generated state, or a meson-baryon molecular state. Another fascinating finding in the coupled-channel framework is the two-pole structure, i.e., the single $\Lambda(1405)$ four-star resonance listed in the Review of Particle Physics (RPP) in versions up to 2018~\cite{ParticleDataGroup:2018ovx}, actually corresponds to two distinct poles on the unphysical Riemann sheet of the complex energy plane~\cite{Oller:2000fj,Jido:2002yz,Garcia-Recio:2002yxy,Hyodo:2011ur,Borasoy:2005ie,Ikeda:2012au,Guo:2023wes}, which are now listed as two items $\Lambda(1380)$ and $\Lambda(1405)$, respectively, in the latest version of RPP~\cite{ParticleDataGroup:2024}. The two pole positions were updated in Ref.~\cite{Lu:2022hwm}   using the state-of-the-art next-to-next-to-leading order chiral potential. We refer to Refs.~\cite{ParticleDataGroup:2024,Hyodo:2020czb,Mai:2020ltx,Meissner:2020khl} for recent reviews about the $\Lambda(1405)$ state.

The two-pole structure of the $\Lambda(1405)$ can be well understood by means of the SU(3) flavor symmetry and group theory~\cite{Jido:2003cb}. For the scattering of Nambu-Goldstone (NG) bosons off the ground-state light baryons, the Weinberg-Tomozawa (WT) term~\cite{Oset:1997it,Oset:2001cn} 
gives the most important contribution at low energies. 
The WT potential projected to the $S$-wave is given by
\begin{eqnarray}
     V_{ij}^{\text{WT}}(\sqrt{s})=-\frac{C_{ij}}{4f^2}(2\sqrt{s}-M_i-M_j)\mathcal{N}_{i}\mathcal{N}_{j},
\end{eqnarray}
with the normalization factor $\mathcal{N}_{i}=\sqrt{(M_i+E_i)/2M_i}$,
where $M_i$ and $E_i$ are the baryon mass and energy in channel $i$, and $f$ is the NG boson decay constant. The $C_{ij}$ coefficients reflect the interaction strength of pertinent channels, which are calculated from the leading order (LO) meson-baryon chiral Lagrangian~\cite{Oset:1997it}. For the $I=0$ and $S=-1$ sector of interest here, there are four coupled channels: $\pi\Sigma$, $\Bar{K}N$, $\eta\Lambda$, and $K\Xi$.  
In the SU(3) limit, the product of the NG boson octet and the light baryon octet can be decomposed into irreducible representations. The decomposition reads
\begin{eqnarray}
	 \bf{8}\otimes \bf{8}=\bf{1}\oplus \bf{8}_s \oplus \bf{8}_a\oplus \bf{10}\oplus \bf{\overline{10}}\oplus \bf{27},
\end{eqnarray}
where the subscripts ``s" and ``a" refer to the symmetric and antisymmetric representations, respectively~\cite{Jido:2003cb}. The WT potential can also be given in a basis of the SU(3) states, and the corresponding coefficient matrix is then diagonalized as
\begin{eqnarray}
C_{\alpha\beta}^{\text{SU(3)}} = \sum_{i,j} \mathcal{D}_{\alpha i} C_{ij} \mathcal{D}_{\beta j}  = \text{diag}(6, 3, 3, 0, 0, -2 ),
\label{eq:cg}
\end{eqnarray}
where $\mathcal{D}_{\alpha i}$ and $\mathcal{D}_{\beta j}$ represent the SU(3) Clebsch-Gordan coefficients and the index $\alpha$($\beta$) runs over the $\bf{1}$, $\bf{8}_s$, $\bf{8}_a$, $\bf{10}$, $\bf{\overline{10}}$, and $\bf{27}$ representations~\cite{Jido:2003cb,deSwart:1963pdg}. The crucial observation from the above equation is that the meson-baryon interaction is attractive in the singlet and octet representations, with the corresponding coefficients being positive, and correspondingly one pole is dynamically generated in each of these three representations. 
This coupled-channel chiral dynamics~\cite{Jido:2003cb,Xie:2023cej,Xie:2023jve} is the origin of the two-pole structure. The $\Lambda(1380)$ and $\Lambda(1405)$ are found to be connected to the dynamically generated singlet and octet ($\bf{8}_s$) poles, respectively; the assignment could be interchanged at the next-to-leading order (NLO)~\cite{Guo:2023wes} (see Ref.~\cite{Zhuang:2024udv} for an alternative analysis). 
The degeneracy of the two octets in the SU(3) limit in Eq.~\eqref{eq:cg} is broken at NLO~\cite{Guo:2023wes}. Using physical masses of the baryons and mesons, the $\bf{8}_a$ octet pole evolves into the $\Lambda(1680)$ resonance.

Besides the $\Lambda(1405)$, there are some other candidates which also have two-pole structures, such as $D_0^*(2300)$, $D_1(2430)$~\cite{Kolomeitsev:2003ac,Guo:2006fu,Guo:2006rp,Albaladejo:2016lbb} and so on. For instance, the $D_0^*(2300)$, similar to the $\Lambda(1405)$, is supposed to be dynamically generated from the NG boson octet scattering off the ground state charmed meson triplet. As a result, not one single pole, but one sextet and one anti-triplet $D_0^*$ poles exist in the complex energy plane~\cite{Albaladejo:2016lbb}. As stressed in, e.g., Refs.~\cite{Xie:2023cej,Xie:2023jve}, the emergence of two-pole structures is very common in the coupled-channel chiral dynamics. Since the positions of the two poles are usually close to each other, one may only see a single effective resonant structure in pertinent invariant mass spectrum. 
Subsequently, one important question is how the two poles living in the same channels can be unambiguously identified in experiments. 
For the two-pole structure of $\Lambda(1405)$, considering that one pole couples stronger to the $\pi\Sigma$ channel, while the other couples stronger to the $\Bar{K}N$ channel, the authors in Ref.~\cite{Jido:2003cb} suggested that it may be possible to find out the existence of the two resonances by performing different experiments. However, this two-pole structure has still not been satisfactorily experimentally resolved up to now. We refer to Ref.~\cite{Meissner:2020khl} for a concise review concerning the two-pole structure phenomena. In this Letter, we propose a novel method to 
identify the two-pole structures using an $\text{SU(3)}$ flavor filter.

{\it Heavy quarkonium decays.} The predicted masses of both $\Lambda(1380)$ and $\Lambda(1405)$ are lower than the $\bar{K}N$ threshold, therefore one can only observe both resonances simultaneously in the $\pi\Sigma$ spectrum. We consider the three-body decay process $Y\to \bar{\Lambda}\Sigma\pi$, where $Y$ represents a charmonium or bottomonium state with proper quantum numbers. If the final state $\pi\Sigma$ is produced from an intermediate hyperon decay,  since $Y$ is an SU(3) singlet while the  $\bar{\Lambda}$ is an SU(3) flavor octet member,  the intermediate state producing the final $\pi\Sigma$ has to belong to an SU(3) octet in the SU(3) symmetric limit. Likewise, for the process $Y\to \bar{\Lambda}(1520)\Sigma\pi$, since $\bar{\Lambda}(1520)$ is generally supposed to be an SU(3) singlet with $J^P=3/2^-$~\cite{Zhang:2013sva,Kamano:2015hxa,Matveev:2019igl,Sarantsev:2019xxm,Klempt:2020bdu}, the intermediate state producing the final $\pi\Sigma$ has to belong to an SU(3) singlet in this case. Consequently, it is possible to single out the flavor octet and singlet poles in the $Y\to \bar{\Lambda}\Sigma\pi$ and  $Y\to \bar{\Lambda}(1520)\Sigma\pi$ process, respectively. Considering that the SU(3) flavor symmetry is an approximate symmetry in quantum chromodynamics (QCD) with corrections about $m_s/\Lambda_\text{QCD}\approx 30\%$, where $m_s$ is the strange quark mass and $\Lambda_\text{QCD}$ is the nonperturbative QCD scale, the above proposed filter should work well.


Such a flavor filter is feasible at currently running experiments. For instance, the BESIII experiment has collected around 10 billion $J/\psi$ events and 3 billion $\psi(2S)$ events~\cite{BESIII:2020nme}. In the future Super Tau-Charm Facility, $3.4\times 10^{12}$ $J/\psi$ or $6.4\times 10^{11}$ $\psi(2S)$ events per year are expected~\cite{Achasov:2023gey}. 
Furthermore, the branching fractions of relevant decay channels to be discussed here are quite sizable~\cite{ParticleDataGroup:2022pth}, and thus a clear identification of the two-pole structure of the $\Lambda(1405)$ can be expected.

\begin{figure}[tb]
	\centering
	\includegraphics[width=1.0\linewidth]{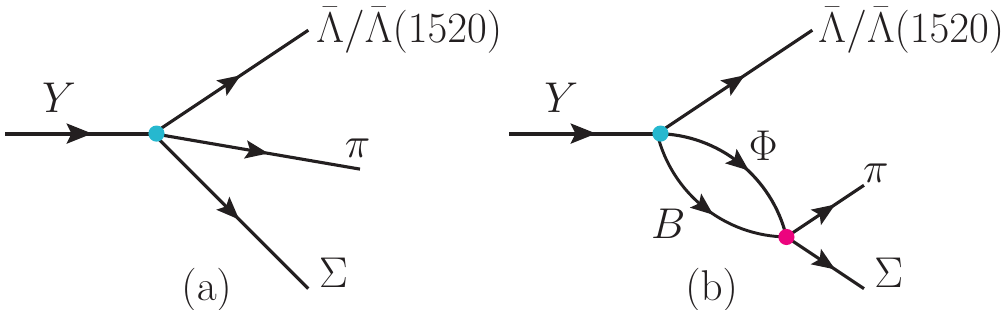}
	\caption{Diagrams for the heavy quarkonium state $Y$ decaying into $\bar{\Lambda}\Sigma\pi$ or $\bar{\Lambda}(1520)\Sigma\pi$. Here, $\Phi$ and $B$ represent the intermediate NG boson and light baryon, respectively.}
	\label{fig:feynmann}
\end{figure}

{\it Formalism and results.}
A diagrammatic description for the $Y\to \Bar{\Lambda}\Sigma\pi$ and $Y\to \Bar{\Lambda}(1520)\Sigma\pi$ decays is presented in Fig.~\ref{fig:feynmann}, where the three light hadrons are first produced via a short-distance process as shown in diagram~(a), which can then be followed by a meson-baryon final-state interaction (FSI) as illustrated in diagram~(b). The meson-baryon FSI will be treated using the chiral unitary approach, and the two $\Lambda(1405)$ poles will be dynamically generated in the coupled-channel chiral dynamics. 
The $\chi_{c0}$ decays into $\bar \Sigma\Sigma\pi$ and $\bar \Lambda\Sigma\pi$ have been studied in Refs.~\cite{Wang:2015qta,Liu:2017hdx}. However, the flavor filter to identify the two-pole structure of $\Lambda(1405)$ has not been discussed.

The short-distance decay vertex of $J/\psi$ or $\psi(2S)$ into $\bar{B}B\Phi$ may be parametrized by the effective Lagrangian
\begin{eqnarray}\label{eq:LagJpsi}
    \mathcal{L}_\psi^\text{I}=\Tilde{D}  \left\langle \bar{B} \gamma_\mu \gamma_5 \{ \Phi, B  \} \right\rangle \psi^\mu +\Tilde{F}  \left\langle \bar{B} \gamma_\mu \gamma_5 [ \Phi, B  ] \right\rangle \psi^\mu,
\end{eqnarray}
where $\Phi$ and $B$ represent the ordinary SU(3) matrices for the pseudoscalar meson octet and  light baryon octet, respectively, and the symbol $\langle\cdots \rangle$ stands for the trace in the flavor space. The coupling constants $\Tilde{D}$ and $\Tilde{F}$ can be determined from reproducing experimental measurements. For the vector charmonia decays in this work, we use the PDG branching ratios~\cite{ParticleDataGroup:2022pth} of four channels $\bar{\Lambda}\Sigma\pi$, $\bar{\Lambda}N\bar{K}$, $\bar{\Lambda}\Lambda\eta$, and $\bar{\Sigma}N\bar{K}$ to fit the coupling constants in Eq.~(\ref{eq:LagJpsi}). 
The $\pi\Sigma$ line-shape depends on the ratio $\mathcal{R}_{\Tilde{F}/\Tilde{D}}$ of the two coupling constants.  For the $J/\psi$ and $\psi(2S)$ decays, the fit gives $\mathcal{R}_{\Tilde{F}/\Tilde{D}}=0.18\pm 0.03$ and $\mathcal{R}_{\Tilde{F}/\Tilde{D}}=0.50\pm 0.06$, respectively.

For the vector charmonia decaying into $\bar{\Lambda}(1520) B\Phi$, the effective Lagrangian can be constructed as
\begin{eqnarray}\label{eq:Lag1520}
    \mathcal{L}_\psi^{\rm II}=g_0  \bar{\Lambda}_\mu \gamma_5  \left\langle  \Phi B  \right\rangle \psi^\mu,
\end{eqnarray}
where the Rarita-Schwinger field $\Lambda_\mu$ represents the SU(3) singlet spin-3/2 hyperon $\Lambda(1520)$, and $g_0$ is the coupling constant. This are currently no data concerning $J/\psi$ and $\psi(2S)$ decaying into $\bar{\Lambda}(1520) B\Phi$. However, since only one single coupling $g_0$ appears in $\mathcal{L}_\psi^{\rm II}$, the $\pi\Sigma$ line-shape will not be affected by the $g_0$ value.

Taking into account the meson-baryon FSI using the chiral unitary approach, the amplitude for the decay of a charmonium into channel $i$ can be written as 
\begin{eqnarray}\label{eq:fullamplitude}
    t_i=\Tilde{V}_i +\sum_j \Tilde{V}_j G_j T_{ji},
    \label{eq:Adecay}
\end{eqnarray}
where $\Tilde{V}_i$ corresponds to the short-distance decay amplitude shown in Fig.~\ref{fig:feynmann}~(a) and can be derived from Eqs.~(\ref{eq:LagJpsi}) and (\ref{eq:Lag1520}). The index $j$ runs over four coupled channels: $\pi\Sigma$, $\Bar{K}N$, $\eta\Lambda$, and $K\Xi$. According to the Lagrangian $\mathcal{L}_\psi^\text{I}$, for $\psi\to \bar{\Lambda}B\Phi $, the effective coupling constant $h_j$ for each channel reads
\begin{eqnarray}
&& h_{\pi\Sigma}=-\sqrt{2}\Tilde{D},\ \ \ h_{\bar{K}N}=-\sqrt{\frac{1}{3}}\Tilde{D}-\sqrt{3}\Tilde{F}, \nonumber \\
&& h_{\eta\Lambda}=-\sqrt{\frac{2}{3}}\Tilde{D},\ \ \ h_{K\Xi}=\sqrt{\frac{1}{3}}\Tilde{D}-\sqrt{3}\Tilde{F}.
\end{eqnarray}
For $\psi\to \bar{\Lambda}(1520)B\Phi $, the effective couplings read
\begin{eqnarray}
&& h_{\pi\Sigma}=-\sqrt{3}g_0,\ \ \ h_{\bar{K}N}=\sqrt{2}g_0, \nonumber \\
&& h_{\eta\Lambda}=g_0,\ \ \ h_{K\Xi}=-\sqrt{2}g_0.
\end{eqnarray}

The unitarized scattering amplitude $T_{ij}$ takes the form
\begin{eqnarray}
    T_{ij}=V_{ij}+V_{ik}G_k T_{kj}.
\end{eqnarray}
The loop function $G_k$ can be regularized using the dimensional regularization. For the LO WT interaction, we adopt the same subtraction constants as those in Refs.~\cite{Jido:2003cb,Oset:2001cn}, i.e., $a_{\pi\Sigma}=-2.00$, $a_{\bar{K}N}=-1.84$, $a_{\eta \Lambda}=-2.25$, and $a_{K \Xi}=-2.67$, and use them for the loop function in Eq.~\eqref{eq:Adecay} as well. With these parameters, one finds two poles located in the vicinity of 1.4~GeV: Pole 1 at $(1390-i66)$ MeV and pole 2 at $(1426-i16)$ MeV~\cite{Jido:2003cb}. 
The physical masses of pertinent mesons and baryons are used to obtain the above predictions, which implies that the SU(3) symmetry is already broken. 
The analysis in Ref.~\cite{Jido:2003cb} shows that 
pole 1 and pole 2 develop from the singlet and octet ($\bf{8}_s$) poles, respectively~\cite{Jido:2003cb}.
For clarity, we name pole 1 and pole 2 as $\Lambda(1380)$ and $\Lambda(1405)$ in the following discussion, although their positions vary a bit in different approaches. 
Since SU(3) is a good approximate symmetry, we expect that in the physical situation the main components of these poles are still SU(3) singlet and octet, respectively.

\begin{figure}[tb]
	\centering
	\includegraphics[width=0.8\linewidth]{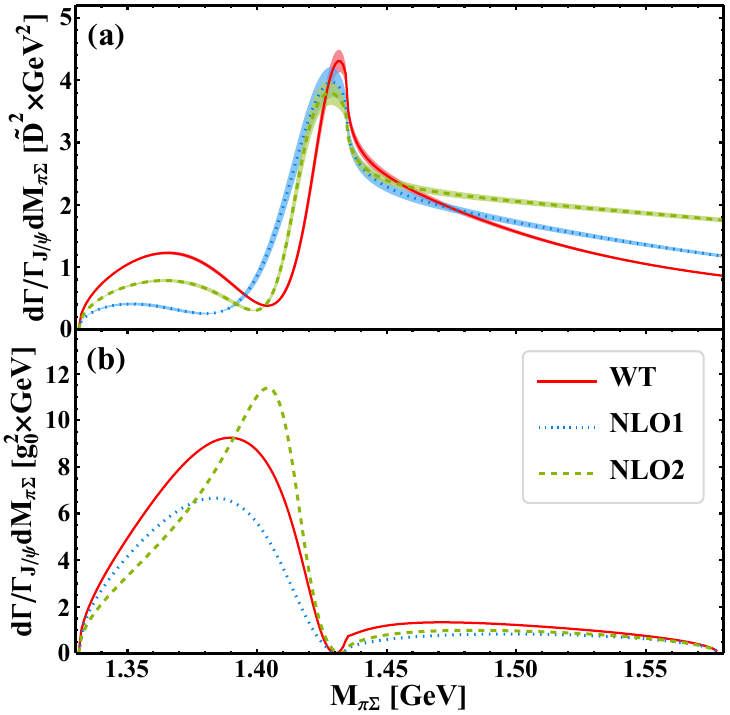}
	\caption{ $\pi\Sigma$ invariant mass distributions in (a) $J/\psi\to \Bar{\Lambda}\Sigma\pi$ and (b) $J/\psi\to \Bar{\Lambda}(1520)\Sigma\pi$. The narrow bands in (a) are obtained by considering the uncertainty in the ratio $\mathcal{R}_{\Tilde{F}/\Tilde{D}}$. }
	\label{NLO-Jpsi-Lambda}
\end{figure}

The $\pi\Sigma$ invariant mass spectra for the two $J/\psi$ decay modes are displayed in Fig.~\ref{NLO-Jpsi-Lambda}. Firstly we focus on the LO results (solid lines). From Fig.~\ref{NLO-Jpsi-Lambda}~(a), it can be seen that the narrower octet pole $\Lambda(1405)$ is much more prominent compared with the broader singlet pole $\Lambda(1380)$. This implies that the $\Lambda(1405)$ state can be effectively singled out in the $J/\psi\to \Bar{\Lambda}\Sigma\pi$ decay.
On the other hand, for the $J/\psi\to \Bar{\Lambda}(1520)\Sigma\pi$ decay, as expected, the SU(3) singlet pole $\Lambda(1380)$ is much more prominent compared with the octet pole, as can be seen from Fig.~\ref{NLO-Jpsi-Lambda}~(b).
As a result, we can effectively single out the singlet $\Lambda(1380)$ state in such a process.

Moreover, the signal for the octet pole $\Lambda(1405)$ is shielded beneath a dip around the $N\bar K$ threshold. This is a consequence of that the universal mechanism for the emergence of a dip proposed in Ref.~\cite{Dong:2020hxe}: 
For the $\Lambda(1520)$ being an SU(3) singlet, the $B\Phi$ pair produced from the $J/\psi\to \bar{\Lambda}(1520)B\Phi$ must be largely an SU(3) singlet, and the octet enters only as an intermediate state via SU(3) breaking; in that case, 
the strong $S$-wave attraction between $N\bar K$ in the octet naturally leads to a dip (for a recent application of the same mechanism in explaining the $X(3872)$ as a dip in the cross section of $e^+e^-\to J/\psi \pi^+\pi^-$, we refer to Ref.~\cite{Baru:2024ptl}).

It should be mentioned that the other octet pole $\Lambda(1680)$ is far from the $\Lambda(1380)$ and $\Lambda(1405)$ and will not pollute the two-pole structure we are interested in here.

The above results were obtained with the LO WT interaction. Next, we further consider the NLO contributions for the NG bosons scattering off the light baryons. There have been various NLO calculations in the literature~\cite{Borasoy:2005ie,Ikeda:2011pi,Ikeda:2012au,Guo:2012vv,Mai:2014xna,Guo:2023wes}. Two sets of parameters from Ref.~\cite{Ikeda:2012au} and Ref.~\cite{Guo:2023wes} are employed in our calculation. In Ref.~\cite{Ikeda:2012au}, the pole positions of the $\Lambda(1380)$ and $\Lambda(1405)$ in the full NLO scheme (NLO1) were found as $(1381-i81)$ MeV and $(1424-i26)$ MeV, respectively.
In Ref.~\cite{Guo:2023wes}, the pole positions at NLO without the Born terms (NLO2) were found as $(1415-i165.7)$ MeV and $(1417.9-i15.6)$ MeV. The NLO chiral Lagrangian can be found explicitly in Ref.~\cite{Borasoy:2005ie}. Both the NLO contact and Born terms break the SU(3) symmetry.
The numerical NLO results in the two schemes NLO1 and NLO2 are displayed in Fig.~\ref{NLO-Jpsi-Lambda}. One may notice that although the pole positions vary in different schemes, the line-shape behaviors in the NLO schemes remain similar as the LO ones. 
These numerical results show the reliability of the flavor filter method against NLO corrections.

\begin{figure}[tb]
	\centering
	\includegraphics[width=0.8\linewidth]{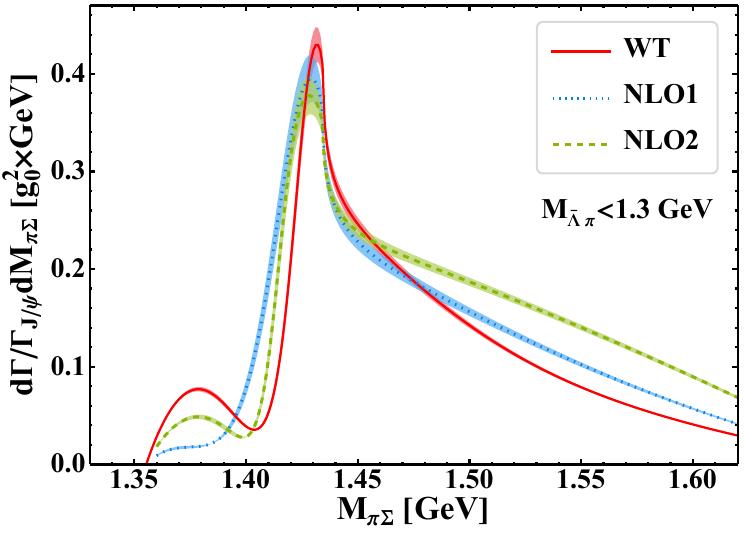}
	\caption{$\pi\Sigma$ invariant mass distribution  in $J/\psi\to \Bar{\Lambda}\Sigma\pi$ with a cut of $M_{\bar{\Lambda}\pi}< 1.3$ GeV.}
	\label{NLO-Jpsi-cut}
\end{figure}

For the process $J/\psi\to \Bar{\Lambda}\Sigma\pi$, in the above discussion we only consider the situation where $\pi\Sigma$ is produced from  intermediate hyperons with $I=0$. As a three-body decay process, it can also receive contributions from the cascade decay $J/\psi\to \Sigma \bar{\Sigma}^{**} \to {\Sigma}\Bar{\Lambda}\pi$, with $\bar{\Sigma}^{**}$ being an intermediate isovector anti-hyperon, such as the $\bar{\Sigma}(1385)$ state. As can be seen from the Dalitz plots of the recent BESIII measurement~\cite{BESIII:2023syz}, the $\bar{\Sigma}(1385)$ signal is sizeable. Although the $\bar{\Sigma}^{**}$ peak appears in the $\pi\bar{\Lambda}$ spectrum, the reflection effect may still pollute the identification of the excited $\Lambda$ state in the $\pi\Sigma$ spectrum. The influence of reflection effects can be eliminated by a proper energy cut in the Dalitz plot. In Fig.~\ref{NLO-Jpsi-cut}, we show the $\pi\Sigma$ distribution with $M_{\bar{\Lambda}\pi}< 1.3$~GeV which cuts out all $\bar \Sigma$ resonances. One notices that the $\pi\Sigma$ line-shape  remains consistent with that in Fig.~\ref{NLO-Jpsi-Lambda}~(a). 
It looks like that the Dalitz plot distribution reported by BESIII in Ref.~\cite{BESIII:2023syz} in the $M_{\bar{\Lambda}\pi}< 1.3$~GeV region has an accumulation of data around $M_{\Sigma^+\pi^-}^2\approx 2$~GeV$^2$, corresponding to the peaking position in Fig.~\ref{NLO-Jpsi-cut},  more than in its neighborhood.
However, a solid conclusion can only be made with an analysis of the Dalitz plot distribution with the advocated cut, which we strongly call for.

\begin{figure}[tb]
	\centering
	\includegraphics[width=0.98\linewidth]{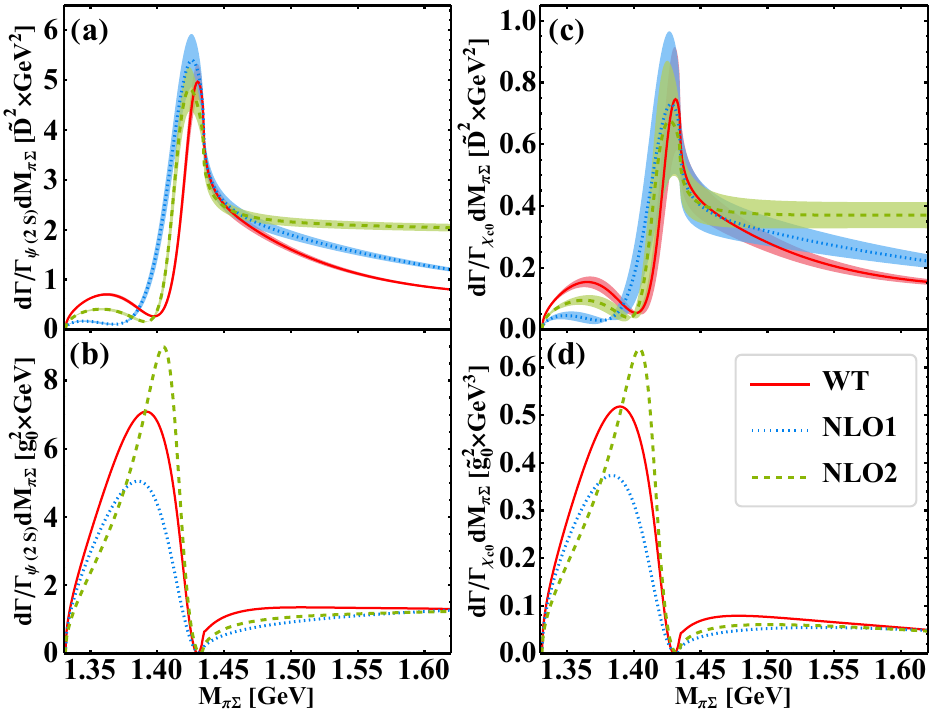}
	\caption{$\pi\Sigma$ invariant mass distributions in (a) $\psi(2S)\to \Bar{\Lambda}\Sigma\pi$, (b) $\psi(2S)\to \Bar{\Lambda}(1520)\Sigma\pi$, (c) $\chi_{c0}\to \Bar{\Lambda}\Sigma\pi$ and (d) $\chi_{c0}\to \Bar{\Lambda}(1520)\Sigma\pi$. The bands in (a) and (c) are obtained by considering the uncertainty in the ratio $\mathcal{R}_{\Tilde{F}/\Tilde{D}}$. 
	}
	\label{Psi2S-chic0}
\end{figure}

In addition to the $J/\psi$ decays, the same idea applies to the $\psi(2S)$ and $\chi_{c0}$ decays into $\Bar{\Lambda}\Sigma\pi$ and $\Bar{\Lambda}(1520)\Sigma\pi$. 

The effective Lagrangian for the $\chi_{c0}$ decaying into $\bar{B}B\phi$ takes the form
\begin{eqnarray}\label{eq:Lagchic0}
	\mathcal{L}_{\chi_{c0}}^{\rm I}=\Tilde{D}_\chi  \left\langle \bar{B}  \gamma_5 \{ \Phi, B  \} \right\rangle \chi_{c0} +\Tilde{F}_\chi  \left\langle \bar{B}  \gamma_5 [ \Phi, B  ] \right\rangle \chi_{c0}.
\end{eqnarray}
For $\chi_{c0}\to \Bar{\Lambda}(1520)B\Phi$, the Lagrangian is given by
\begin{eqnarray}\label{eq:LagChic0-1520}
	\mathcal{L}_{\chi_{c0}}^{\rm II}=i\Tilde{g}_0  \bar{\Lambda}_\mu  \gamma_5 \left\langle  \partial^\mu\Phi B +\Phi \partial^\mu B \right\rangle \chi_{c0}.
\end{eqnarray}
Likewise, the coupling constants $\Tilde{D}_\chi$ and $\Tilde{F}_\chi$ in Eq.~(\ref{eq:Lagchic0}) can be determined from the branching fractions of the $\chi_{c0}$ decays into the $\bar{\Lambda}\Sigma\pi$, $\bar{\Lambda}N\bar{K}$, and $\bar{\Sigma}N\bar{K}$ channels, and we obtain the ratio ${\Tilde{F}_\chi/\Tilde{D}_\chi}=0.29\pm 0.17$.

For the $\psi(2S)$ and $\chi_{c0}$ decays,  
the $\pi\Sigma$ invariant mass distribution curves are displayed in Figs.~\ref{Psi2S-chic0}~(a, b) and (c, d), respectively. As before,  the octet $\Lambda(1405)$ pole is much more significant in the $\Bar{\Lambda}\Sigma\pi$ final states, while the singlet $\Lambda(1380)$ pole dominates  the $\Bar{\Lambda}(1520)\Sigma\pi$ final state.

{\it Conclusion and outlook.}
In summary, we proposed a novel method to unambiguously identify the two-pole structure of $\Lambda(1405)$. The key point of this method is that the two poles are rooted in different irreducible SU(3) flavor representations (one singlet and one octet). 
For a heavy quarkonium state $Y$ decaying into $\bar{\Lambda}\Sigma\pi$ and $\bar{\Lambda}(1520)\Sigma\pi$, the octet  and singlet poles can be singled out separately. This SU(3) flavor filter works even considering the flavor symmetry breaking. 
Although the positions of the two poles vary depending on the parameters used in the literature, their production weights in the $\bar{\Lambda}\Sigma\pi$ and $\bar{\Lambda}(1520)\Sigma\pi$ channels are rather different, and the line-shape behaviors remain relatively stable using different parameter sets.

This flavor filter is experimentally accessible, since running experiments, e.g. BESIII, have already accumulated huge data samples. 
With the already available data in Ref.~\cite{BESIII:2023syz}, a solid conclusion about the possible resonant structures in the $\pi\Sigma$ distribution can be made once the cut $M_{\bar\Lambda\pi}$ is made.
It is therefore promising to solve the long-standing  $\Lambda(1405)$ puzzle by employing this flavor filter. 
The present proposal, if confirmed, will provide the first highly nontrivial experimental confirmation on the existence of the two-pole structure of the $\Lambda(1405)$, complementary to the recent lattice QCD~\cite{BaryonScatteringBaSc:2023zvt,BaryonScatteringBaSc:2023ori} and chiral effective field theory studies~\cite{Xie:2023cej,Xie:2023jve,Guo:2023wes,Ren:2024frr,Zhuang:2024udv}.

The same flavor filter idea can also be used to identify other multiple-pole structures in coupled-channel dynamics. For instance, in the decay of a bottomonium (such as $\Upsilon$) into $\bar{D}^* D\pi$, the anti-triplet pole of $D_0^*(2300)$ may be singled out, since the bottomonium is an SU(3) singlet and $\bar{D}^*$ belongs to a triplet.
Note that in this case, the initial state cannot be in the charmonium mass region since the nature of charmonium(-like) resonances which can decay into $\bar D^* D\pi$ is under debate and they might not be ideal SU(3) singlets~\cite{Chen:2019mgp}.

\begin{acknowledgments}

This work is partly supported by the National Key R\&D Program of China under Grant No.~2023YFA1606703; by the National Natural Science Foundation of China under Grants No.~11975165, No.~12235018, No.~12075288, No.~12125507, No.~12361141819, and No.~12047503; by the Chinese Academy of Sciences (CAS) under Grants No.~YSBR-101 and No.~XDB34030000. It is also supported by the Youth Innovation Promotion Association of CAS.

\end{acknowledgments}

	

%

\end{document}